\documentstyle[amsfonts,prl,multicol,aps]{revtex}
\input{psfig.sty}

\begin{document}
\twocolumn[
\hsize\textwidth\columnwidth\hsize\csname @twocolumnfalse\endcsname

\title{Correlations in optical phonon spectra of complex solids}
\author{G. Fagas $^{1}$, Vladimir I. Fal'ko $^{1}$, C. J. Lambert $^{1}$, and Yuval
Gefen $^{2}$}
\address{$^{1}$ Department of Physics, Lancaster University, Lancaster LA1 4YB, United Kingdom\\
$^{2}$ Department of Condensed Matter Physics,
Weizmann Institute of Science, 76100 Rehovot, Israel}
\maketitle

\begin{abstract}
Spectral correlations in the optical phonon spectrum of a solid with a
complex unit cell are analysed using the Wigner-Dyson statistical approach.
Despite the fact that all force constants are real, we find that the statistics
are predominantly of the GUE type depending on the
location within the Brillouin zone of a crystal and the unit cell symmetry.
Analytic and numerical results for the crossover from GOE to GUE statistics
are presented.
\end{abstract}
]

Wigner-Dyson statistical analysis \cite{Mehta} has become a
widespread approach to characterise quantum spectra of complex dynamical
systems, such as nuclei, atoms and molecules, disordered quantum electron
systems and electromagnetic resonances in chaotic cavities \cite
{Mehta,Simons,Guhr,Berry,Rsymm}. Both experimental and
theoretical studies of such systems point to the fact that
the quantum energy levels
universally obey the same mutual correlations and spectral
rigidity as the eigenvalues of random Gaussian-distributed matrices. In the
present paper, we apply such a statistical approach to characterise
vibrational spectra of multi-component solids with a complex unit cell
structure, in order to assess the correlation properties of their optical
phonon spectra taken at various positions in the Brillouin zone (BZ).

In random matrix theory (RMT), different ensembles of matrices,
which are defined by their fundamental symmetry, result in
distinct level correlation properties. The two most common
in condensed matter physics are the Gaussian ensemble of
real symmetric matrices (GOE) which adequately describes the
spectral correlations in
quantum chaotic electron systems with time reversal symmetry, and the
Gaussian ensemble of Hermitian matrices (GUE) which describes
chaotic electron billiards with time-reversal symmetry broken by a magnetic
field. One may expect a set of $3N$ coupled classical oscillators to be
an example of a system with time-reversal symmetry, thus obeying
the GOE spectral statistics. This is indeed the case with the
acoustic spectroscopy of irregularly shaped solid resonators
\cite{Ellegaard,Leitner}, or with the spectrum of a regularly shaped system
consisting of coupled oscillators whose masses are random
\cite{Fagas}. Below we demonstrate the counter-intuitive result that,
on the whole, the spectral correlations
of the optical phonon modes associated with the same (albeit
arbitrary) point of the BZ of a complex solid obey the GUE statistics.
This complies with an earlier observation \cite{Mucciolo} on the electronic
structure of highly excited bands in solids.
In particular, we report a study of the distribution function $P(s)$ of
the nearest-level-spacing in vibrational spectra of a complex solid
based on numerical simulations of crystalline structures,
both with and without mirror reflection symmetry in the unit cell. We also
analyse the dependence of statistical properties on the phonon wave number
${\bf Q}$ within the BZ, and obtain a detailed description of the
crossover between the limiting regimes of GOE-type correlations specific to
the phonon frequencies exactly at the center of the BZ where ${\bf Q}%
=0$, and of GUE-type for sufficiently large values of $Q$.

To simulate a complex solid, we have adopted the following model. The unit cell of
a crystal was taken in the form of a parallelepiped consisting of $%
N=8\times 10\times 12$ atoms with equal pair interactions but randomly
chosen masses arranged on an fcc lattice. The unit cell size $L$
determines the periodicity of the Bravais lattice of the entire solid, $%
{\bf L}=n_{1}{\bf L}_{1}+n_{2}{\bf L}_{2}+n_{3}{\bf L}_{3}$. The spectrum of
optical phonons in it can be found by determining all $3N$ normal modes
corresponding to the linearised set of equations for coupled harmonic
oscillators, $m_{{\bf j}}\ddot{u}_{{\bf j}}^{\alpha }=-\sum_{{\bf iL}}K_{%
{\bf ij}}^{\alpha \beta }({\bf L})\;u_{{\bf i+L}}^{\beta }$,
where ${\bf i},{\bf j}$ are atomic positions within the unit cell
and $\alpha,\beta$ denote cartesian components.
Fourier transforming we obtain for each given
point in the BZ (wave number ${\bf Q}$),
\begin{equation}
K_{{\bf ij}}^{\alpha \beta }({\bf Q})=\sum_{{\bf L}}k(l_{{\bf L+i,j}%
}^{\alpha }l_{{\bf L+i,j}}^{\beta }\;\ e^{\imath {\bf QL}}-4\delta _{{\bf %
L+i,j}}\delta ^{\alpha \beta }),  \label{eq1}
\end{equation}
where $k$ is the interatomic force constant, $l_{{\bf L+i,j}}^{\alpha }={\bf %
(j-i-L)}^{\alpha }/{\bf \left| j-i-L\right| }$, and ${\bf (j-i-L)}$
belongs to the first coordination sphere of an fcc lattice.
Note that the latter implies non-zero matrix elements
only for nearest neighbors and applies restrictions to
the sum over ${\bf L}$ in Eq. (\ref{eq1}).
The sets of phonon frequencies $\{\omega _{k}({\bf Q})\}$
(which result in a spaghetti of $3N$ dispersion curves, each
corresponding to a particular phonon branch) can be obtained by solving
numerically the eigenvalue equation
\begin{equation}
{\rm det}(D-\omega ^{2}\;I)=0\;,\;D({\bf Q})=M^{-1/2}K({\bf Q})M^{-1/2}.
\label{eq2}
\end{equation}
This equation also defines the dynamical matrix, $D({\bf Q})$, in which the
complexity in the composition of a solid is introduced by the random diagonal
matrix $M_{{\bf ij}}^{\alpha \beta }=m_{{\bf i}}\delta ^{\alpha \beta
}\delta _{{\bf ij}}$. For the masses $m_{{\bf j}}$ we have used
a box distribution over the interval of masses $[\langle m \rangle-\delta m,
\langle m \rangle+\delta m]$ with $\delta m / \langle m\rangle =0.3$.
This corresponds to an r.m.s. of mass disorder
$\langle \delta m^{2}\rangle ^{1/2}/\langle m\rangle \approx 0.17$.
The mean value $\langle m\rangle$ defines the cut-off frequency $\omega_c=
\sqrt{8k/\left\langle m\right\rangle }$ of the vibrational spectrum in the
disorder-free limit and sets a characteristic scale to measure the eigenfrequencies, whereas
lengths are measured in units of the lattice constant. As explained below,
such a model also allows us to exploit an analogy between the numerical results
obtained by us and
the properties of spectra of systems exhibiting quantum diffusion. To avoid
complications brought into the problem by the localisation effects related
to the vibrations of light atoms in a heavy matrix, we restrict
our statistical analysis to the frequency range $\omega /\omega_c<0.95$.

At the middle of the BZ (${\bf Q}=0$) and at other symmetry
points, such as ${\bf Q}=(0,\pi /L_{2},\pi /L_{3})$, the
dynamical matrix $D$ is real and symmetric. For an arbitrary ${\bf Q\neq 0}$,
the matrix elements of $D({\bf Q})$ related to the sites on the edges of the
unit cell (we use the nearest-neighbor interaction)
acquire complex phase factors which make the whole dynamical matrix
complex and Hermitian, $D({\bf Q})=D_{S}({\bf Q})+\imath D_{A}({\bf Q})$, where $%
D_{S}({\bf Q})$, $D_{A}({\bf Q})$ are real symmetric and antisymmetric
matrices, respectively. Without imposing any spatial symmetries onto the
unit cell, these two scenarios span the two limiting cases for
the D-matrix symmetry; the difference between these two regimes
manifests itself in the form of the normalized distribution function $P(s)$
of the nearest-level-spacing, $s=(\omega _{k+1}-\omega _{k})/\Delta $, where 
$\Delta $ is the mean level spacing.
\begin{figure}[tbp]
\centerline{\psfig{figure=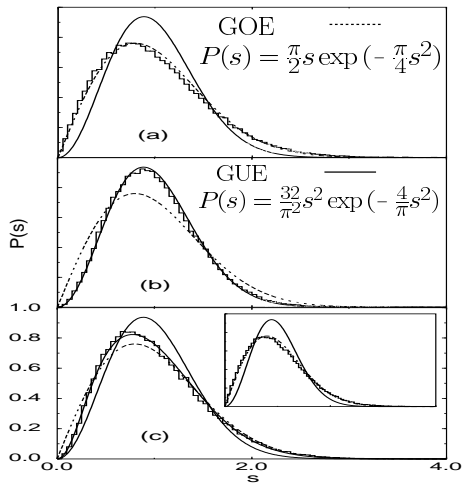,height=7cm,width=7.5cm}}
\caption{Nearest-level-spacing distribution of optical phonon spectra
for a non-symmetric unit cell and various values of ${\bf Q}$:
(a) ${\bf Q}=0$, (b) ${\bf Q}=[0, \protect%
\pi/4L_2, 3\protect\pi/4L_3]$, and (c) ${\bf Q}=[0.05/L_1, 0, 0]$
(inset : ${\bf Q}=[0, \protect\pi/L_2, \protect\pi/%
L_3]$). Fits to histograms are explained in the text.}
\label{zero}
\end{figure}
We have constructed $P(s)$ employing a two-step averaging
procedure. The first step is to use an ensemble of 50 random realizations of
the distribution of masses in the sample. A further averaging of $P(s)$ is
applied over a broad frequency range, namely, $0.45<\omega /\omega_c
<0.95$, for each of the calculated spectra, by using an observation that they
become statistically indistinguishable
after having been rescaled by $\Delta $ \cite{Guhr}.
Note that we do not analyze the lowest part of the spectrum because of poor
statistics. The nearest-level-spacing distribution function for the optical
phonon spectrum at ${\bf Q}=0$ is shown in Fig. \ref{zero}(a). It coincides
with the random matrix theory prediction for real symmetric matrices given
by Wigner-Dyson distribution function for the GOE \cite{Mehta}. In contrast,
optical phonons with ${\bf Q}\not=0$ exhibit a nearest-level-spacing
statistics which is best fitted by the Wigner-Dyson distribution function
for the GUE. A typical $P(s)$-histogram is plotted in Fig. \ref{zero}(b),
and is compared with the GUE analytical result \cite{Mehta}. A similar
observation has been made earlier by Mucciolo {\it et al} \cite{Mucciolo} in
relation to the electronic band structures in crystals. For completeness,
the inset in Fig. \ref{zero}(c) shows the $P(s)$-histogram for a corner
of the BZ, which is of GOE-type. Fig. \ref{zero}(c) also
illustrates the form of $P(s)$ in the intermediate regime between two
distinct statistical classes. It is compared with the result of a fit based on
the RMT prediction \cite{Berry} for an interpolating ensemble between the
GOE and GUE which contains a single fitting parameter.

\begin{figure}[tbp]
\centerline{\psfig{figure=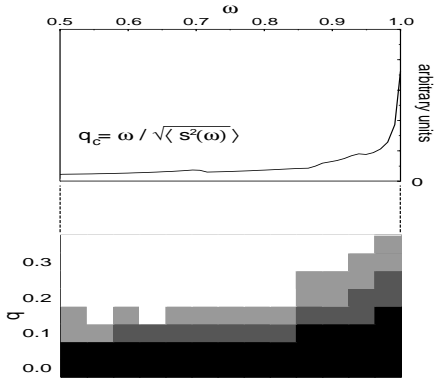,height=6cm,width=6.5cm}}
\caption{GOE-GUE crossover.
The gray-scale plot illustrates the level of deviations of the
calculated $P(s)$ from the GOE result, $\protect\sigma = \frac{\int[%
P(s)-P_{GOE}(s)]^2ds}{\int[P_{GUE}(s)-P_{GOE}(s)]^2ds}$ ($0 < \protect\sigma
< 1$). Black is used for $0 < \protect\sigma < 0.25$, whereas, the width of
the boxes is defined by the spectral window used to built $P(s)$.}
\label{cross}
\end{figure}

The inter-ensemble crossover takes place at relatively small values of ${\bf %
Q}$ \ (the parameter responsible for the imaginary part of the dynamical
matrix), as it happens with a similar crossover in the energy spectra of
chaotic electronic systems in the presence of a weak magnetic field \cite
{Efetov}.  Note also that for small $Q$'s, the function $P(s)$ has the following
asymptotic behavior: it resembles the GUE distribution function
at $s\rightarrow 0$, whereas it follows the GOE analytical result
for large s. This is because the splitting of levels with $s < 1$
is more sensitive to a small antisymmetric addition to a symmetric dynamical
matrix than the splitting of rare pairs with $s \gg 1$. The above fact also implies
that the crossover $P(s)$ has a form which cannot be simply reduced
to a trivial mixing between two typical GOE and GUE distributions \cite{Efetov,VF}.
The crossover studies were based upon
the analysis of spectra of 600 random realizations of atomic masses in the
unit cell for various values of ${\bf Q}$ along ${\bf L}_1$. The result of
numerical analysis of the GOE-GUE crossover in different intervals of
vibrational spectra is shown at the bottom of Fig. \ref{cross} in the form of
a gray scale shading of the parametric plane of $q={\bf Q \cdot L}_1 = QL_1$ and
frequency $\omega $, where a darker colour stresses higher similarity of $%
P(s)$ to that of the orthogonal symmetry class, and white indicates the
dominance of the GUE spectral statistics.

Below we present semiclassical arguments which provide the crossover
value of the rescaled $Q$, $q=QL$, as function of the frequency
of the mode considered. Our approach consists of viewing the dynamics as that
of a wave-packet of lattice vibrations (in our case optical modes), spreading
over a unit cell treated here as a region of disordered medium. This treatment
is analogous to the semiclassical treatment of parametric spectral correlations
developed in the studies of quantum disordered electron systems subjected to a weak
magnetic flux \cite{Simons,Gefen}.
In the present analysis, it is assumed that the participation of each
atom in a given optical phonon mode can be described semiclassically for time
intervals shorter than $t_{H}\sim 1/\Delta $, after which the discreteness
of the spectrum for each value of ${\bf Q}$ starts to dominate.
The spread of vibrations over the unit cell and the role of the individual atoms in
this dynamics is considered as a diffusion of waves through an fcc lattice
with mass disorder and is determined by the interference pattern of a variety of
equally probable diffusion paths \cite{note1}. These paths are independent of the exact value
of ${\bf Q}$, whereas the phases of diffusive waves involved in such an
analysis are large and random, so that one obtains correlated
spectra \cite{Simons,Berry}.

However, the cyclic (albeit non-periodic)
boundary conditions impose
a torus geometry in the unit cell. If $W$ is the total number of
windings around the torus then the type of correlations depends on the phase
factor $e^{i\delta \varphi }$ determined by the phase difference between a path
with a positive W and its time-reversed counterpart
with -W. Therefore, $\delta \varphi$ gives us a
measure of time-reversal symmetry breaking and is controlled by the
'external' parameter $q\ll 1$. A relevant
crossover parameter can then be found by estimating the r.m.s.
value of the symmetry breaking phase $\delta \varphi $, acquired by waves whose
propagation is followed along a diffusive path with the maximal length
allowed by the limit set by the time, $t_{H}\sim 1/\Delta $ \cite{Berry,VF},
where $\Delta \sim 1/L^{3}\nu $ and $\nu(\omega )$ is the acoustic phonon density
of states.
The parameter $q$ affects the phase of the partial amplitudes
only for paths that cross the unit cell, i.e., paths whose 'winding' number
is non-zero. Hence, the total symmetry breaking phase is $\delta \varphi \sim qW$,
where $W=\sum w_{i}$ is a winding number made of approximately $t_{H}/t_{D}$ random
contributions $w_{i}=\pm 1$. The time $t_{D}\sim L^{2}/ D$ with $D$ the
diffusion coefficient, is typical of a diffusive spread over the unit cell of a
vibration with frequency $\omega$. For our model,
$D = \frac{1}{3} \left\langle s^{2}\right\rangle \tau $ where
$\left\langle s^{2}(\omega )\right\rangle $ is the angular and polarization
average of the squared phonon group velocity in the clean limit and
$\tau ^{-1}$ is the scattering rate, caused by the variation
of atomic masses. The latter can be estimated \cite{Tamura} as $\tau ^{-1}({\omega })\sim \omega
^{2}\nu \left( \left\langle \delta m^{2}\right\rangle /\left\langle
m\right\rangle ^{2}\right) $. It follows from our considerations that an estimation
of the r.m.s. value of $\delta \varphi $ yields
$\langle \delta \varphi ^2 \rangle ^{1/2} \sim q \sqrt{t_{H}/t_{D}}$.
The crossover can be assigned to
$\langle \delta \varphi ^2 \rangle ^{1/2} \sim 1$. Therefore, the form of the crossover line is determined
by 
\begin{equation}
q_{c}\sim \sqrt{t_{D}/t_{H}}\sim \sqrt{\Delta L^{2}/\left\langle
s^{2}\right\rangle \tau }\propto \omega /\sqrt{\left\langle s^{2}(\omega
)\right\rangle },  \label{eq3}
\end{equation}
which is shown in the top of Fig. 2
and is in agreement with the numerically obtained gray scale plot.

\begin{figure}[tbp]
\centerline{\psfig{figure=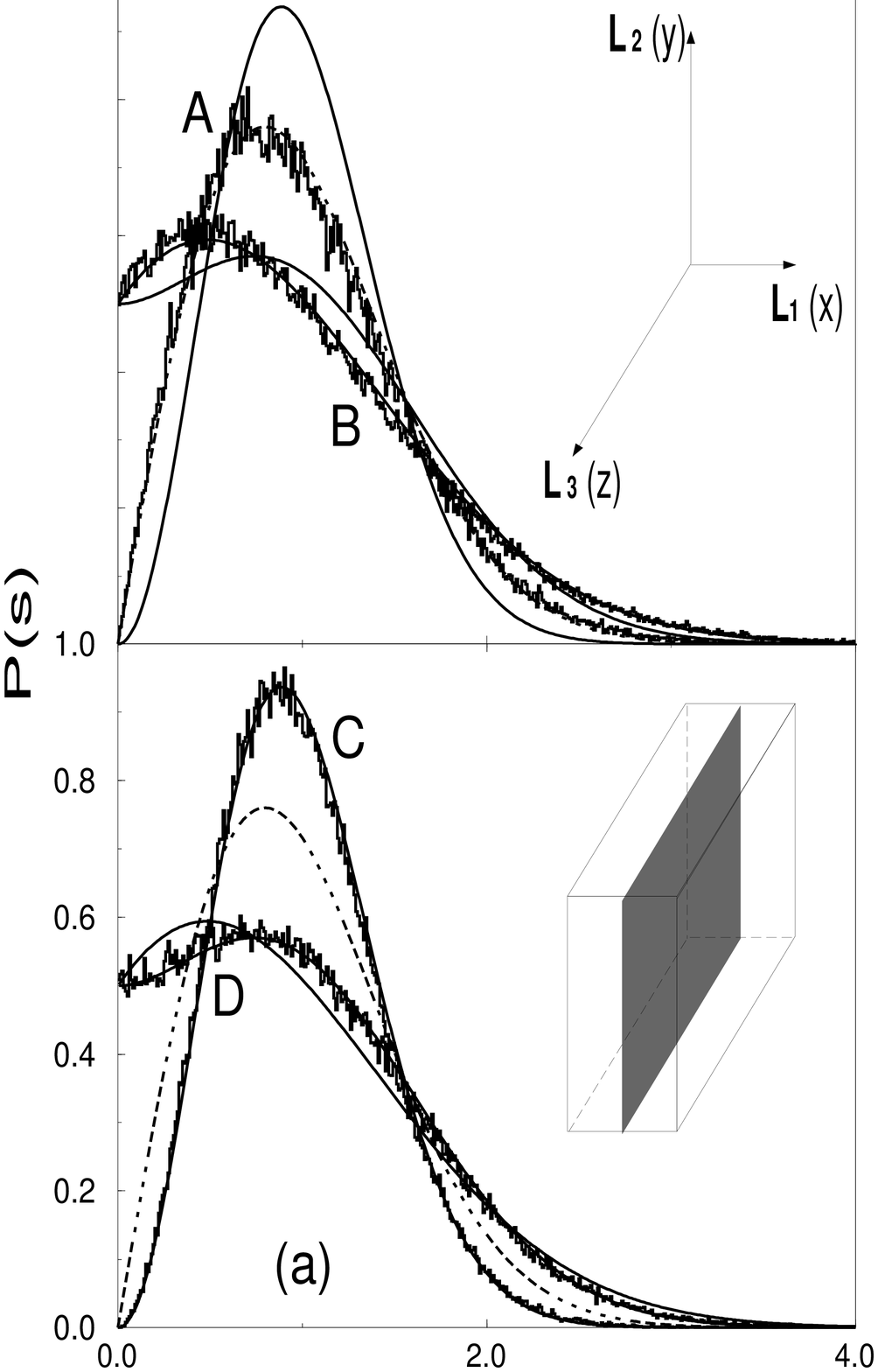,height=5.2cm,width=7.5cm}}
\centerline{\psfig{figure=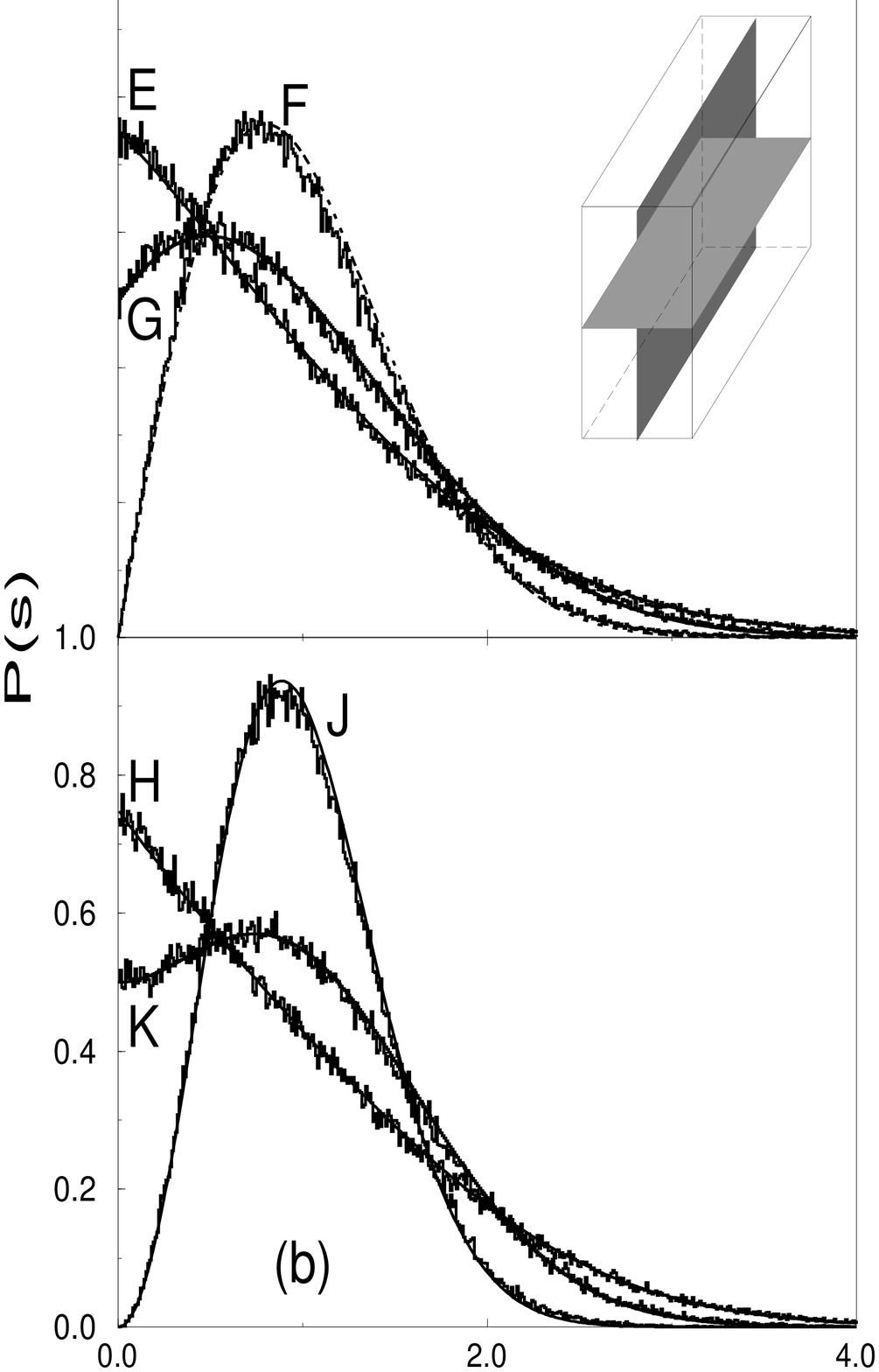,height=5.2cm,width=7.5cm}}
\centerline{\psfig{figure=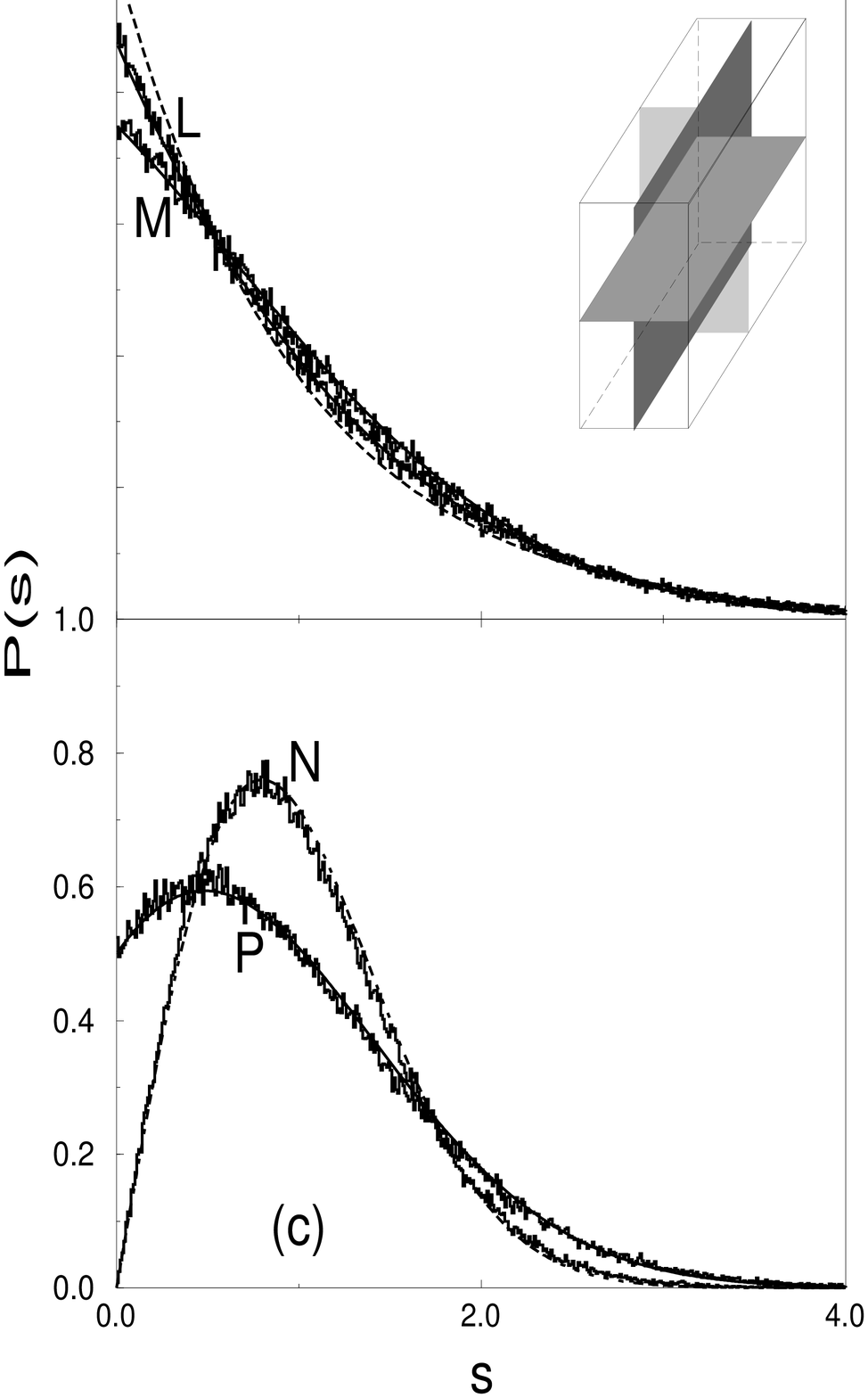,height=5.2cm,width=7.5cm}}
\caption{Nearest-level-spacing distribution of optical phonon spectra
for various unit cell symmetries
and values of ${\bf Q}$:
${\bf Q}=0$ (B, E, L), ${\bf Q}=[\pi/4L_1,0,0]$ (A, G, M),
${\bf Q}=[0,0,\pi/4L_3]$ (D, H), 
${\bf Q}=[\pi/4L_1,3\pi/4L_2,0]$ (C, F, P),
${\bf Q}=[0,3\pi/4L_2,\pi/4L_3]$ (K),
and ${\bf Q}=[3\pi/4L_1,\pi/3L_2,\pi/4L_3]$ (J, N).
Histograms A, F, N(C, J) are compared to the GOE(GUE) analytic result.
Other fits are the $P(s)$ for a number of overlapping
RMT spectra with equal fractional densities:
two GOE(GUE) in B, G, P(D, K), four GOE(GUE) in
E, M(H), and eight GOE(the Poisson distribution is also plotted)
in L.
}
\label{symm}
\end{figure}

The correlation properties of spectra of a solid are sensitive to the
geometrical symmetries \cite{Rsymm} of its unit cell structure, such as the existence of
a mirror plane in it. Below, we extend the numerical analysis to a solid
with $n\leq 3$ mirror symmetry planes in the unit cell, each characterised by unit
vector $\hat{\eta}_{i}$. Numerical simulations similar to the ones described
above (with statistics collected from 100 random realizations of masses
and averaging extended over the spectral window $[0.45,0.95]$) show that the
effect of geometrical symmetries on the spectral statistics depends on the
point in the BZ. This is because the phonon momentum
allows for breaking both the orthogonal and point-group symmetries.
Typical forms of the function $P(s)$ for each symmetry case are shown
in Fig. \ref{symm}, for various wave vectors ${\bf Q}$. The solid lines
in these figures
illustrate the RMT result for overlapping sequences of GOE or GUE spectra
with equal fractional densities \cite{Mehta}. When ${\bf Q}=0$,
the spectra split into sequences of levels corresponding to different
parities, so that $P(s)$ coincides with what one would expect for
$2^{n}$ overlapping GOE's (e.g., $P(s\rightarrow 0)=1-2^{-n}$).

For finite $Q$, spectral statistics is determined by the orientation of
${\bf Q}$ with respect to the mirror symmetry planes,
falling into one of the GOE or GUE classes.
A summary of all distinct statistical limits is given in Table 1.
Note that for ${\bf Q}$ with all non-zero components
perpendicular to a symmetry plane,
correlations are of GOE-type. This effect is due to invariance of
$D({\bf Q})$ under the combination of reflection and complex conjugation
operations, which results in a real representation of the dynamical
matrix \cite{Berry}. Thus, for a unit cell
with 3 mirror symmetry planes only the orthogonal ensemble
statistics is realised for an arbitrary ${\bf Q}$.
The crossover parameter $q_{c}$ which determines which
class from Table 1 should be expected, can be estimated using Eq. (\ref{eq3}).

The authors thank I. Lerner and J. Pendry for discussions. This work was
supported by EPSRC, and by European Union RTN and TMR programmes. Y.G.
acknowledges support by the Centre of Excellence of the Israeli Academy
of Sciences and Humanities.

\begin{table}[tbp]
\begin{tabular}{||c||c|c|c||}
{} & $n=1$ & $n=2$ & $n=3$ \\ \hline\hline
8 $\times$ GOE & - & - &
$\begin{array}{c}
\mbox{symmetry}\\
\mbox{points}
\end{array}$
\\ \hline
4 $\times$ GOE & - &
$\begin{array}{c}
\mbox{symmetry}\\
\mbox{points}
\end{array}$
& ${\bf Q} \cdot \hat{\eta}_i = Q$ \\ \hline
4 $\times$ GUE & - & ${\bf Q} \cdot (\hat{\eta}_i \times \hat{\eta}_j) = Q$ & 
- \\ \hline
2 $\times$ GOE &
$\begin{array}{c}
\mbox{symmetry}\\
\mbox{points}
\end{array}$
&
$\begin{array}{c}
{\bf Q} \cdot \hat{\eta}_i = Q \\
\mbox{and} \\
{\bf Q} \cdot \hat{\eta}_j = 0
\end{array}$
&
$\begin{array}{c}
{\bf Q} \cdot \hat{\eta}_{i(k)} \neq Q \\
\mbox{and} \\
{\bf Q} \cdot \hat{\eta}_j = 0
\end{array}$
\\ \hline
2 $\times$ GUE & ${\bf Q} \cdot \hat{\eta}_i = 0$
&
$\begin{array}{c}
{\bf Q}  \cdot \hat{\eta}_i \neq 0,Q \\
\mbox{and} \\
{\bf Q}  \cdot \hat{\eta}_j = 0
\end{array}$
& - \\ \hline
GOE & ${\bf Q} \cdot \hat{\eta}_i = Q$ &
$\begin{array}{c}
{\bf Q} \cdot (\hat{\eta}_i \times \hat{\eta}_j) = 0 \\
\mbox{and} \\
{\bf Q} \cdot \hat{\eta}_{i(j)} \neq Q
\end{array}$
& any other \\ \hline
GUE & any other & any other & -
\end{tabular}
\caption{Summary of all distinct statistical limits of optical phonon spectra
for
$n\leq 3$ mirror symmetry planes in the unit cell, each characterised by unit
vector $\hat{\eta}_{i}$. RMT ensembles preceded by a number denote a
sequence of corresponding overlapping spectra with equal fractional
densities. Each box contains the condition for these statistics
to be realised at some point ${\bf Q}$ in the BZ.}
\end{table}


\begin{references}
\bibitem{note1}  For a unit cell with no internal structure, as in our model,
the diffusion of wave-packets at the level of the unit cell can be regarded as
propagating acoustic waves across a piece of disordered medium.
\end{references}
\end{document}